\documentclass[doublecol]{epl2}
\usepackage{amsmath}
\definecolor{linkcolor}{rgb}{0,0,0.6} 
\usepackage[pdftex,colorlinks=true,
	pdfstartview = FitV,
	linkcolor    = linkcolor,
	citecolor    = linkcolor,
	urlcolor     = linkcolor,
	hyperindex   = true,
	hyperfigures = false]{hyperref}

\newcommand{\dd}{\text{d}}
\newcommand{\ee}{\text{e}}
\newcommand{\ii}{\text{i}}

\title{Extracting maximum power from active colloidal heat engines}

\author{D. Martin\inst{1,2} \and C. Nardini\inst{1,3} \and M. E. Cates\inst{1} \and \'E. Fodor\inst{1}}
\shortauthor{D. Martin \etal}

\institute{                    
  \inst{1} DAMTP, Centre for Mathematical Sciences, University of Cambridge, Wilberforce Road, Cambridge CB3 0WA, UK
	\\
	\inst{2} Universit\'e Paris Diderot, Sorbonne Paris Cit\'e, MSC, UMR 7057 CNRS, 75205 Paris, France
	\\
  \inst{3} Service de Physique de l'\'Etat Condens\'e, CNRS UMR 3680, CEA-Saclay, 91191 Gif-sur-Yvette, France
}

\pacs{05.70.Ln}{Nonequilibrium and irreversible thermodynamics}
\pacs{05.40.-a}{Fluctuation phenomena, random processes, noise and Brownian motion}
\pacs{82.70.Dd}{Colloids}

\abstract{
	Colloidal heat engines extract power out of a fluctuating bath by manipulating a confined tracer. Considering a self-propelled tracer surrounded by a bath of passive colloids, we optimize the engine performances based on the maximum available power. Our approach relies on an adiabatic mean-field treatment of the bath particles which reduces the many-body description into an effective tracer dynamics. It leads us to reveal that, when operated at constant activity, an engine can only produce less maximum power than its passive counterpart. In contrast, the output power of an isothermal engine, operating with cyclic variations of the self-propulsion without any passive equivalent, exhibits an optimum in terms of confinement and activity. Direct numerical simulations of the microscopic dynamics support the validity of these results even beyond the mean-field regime, with potential relevance to the design of experimental engines.
}

\begin{document}

\maketitle


Colloidal heat engines operate at the microscale by manipulating a colloidal tracer immersed in a fluctuating bath~\cite{Martinez2017}. Some work can be extracted from the bath fluctuations by applying a confining potential to the tracer, whose parameter are varied in time by an external operator~\cite{Seifert2012}. During the past decade, several experimental realizations have demonstrated the feasibility of building such engines, both for an equilibrium and a nonequilibrium bath, which operate with some cyclic protocols inspired by the ones of macroscopic engines~\cite{Blickle2011, Quinto-Su2014, Martinez2016, Sood2016}.

To rationalize the properties of colloidal engines, at variance with macroscopic engines, the tools of standard thermodynamics are not sufficient. Indeed, colloids undergo substantial fluctuations which affect the energy transfers with the bath. The framework of stochastic thermodynamics extends the definition of energetic observables, such as work and heat, in terms of the microscopic stochastic dynamics of the tracer~\cite{Sekimoto1998, Seifert2012}. It provides the appropriate tools to investigate the engine properties for finite-time cycles, thus allowing one to quantify the output power~\cite{Shiraishi2016, Pietzonka2017}. Although the pioneer works were primarily developed for equilibrium systems, some generalisations considering a nonequilibrium medium instead of the thermal bath have been put forward~\cite{Sagawa2014, Kanazawa2014, Zakine2017, Wulfert2017}.

The study of colloidal heat engines is driven by optimizing their performances. One interesting question is how to extract maximum power given the properties of the underlying bath and the details of the driving protocols. Several studies have considered non-cyclic protocols in minimal model systems where quasistatic protocols are known to be optimal~\cite{Jarzynski2008, Aurell2011, Rotskoff2017, Solon2017}. One can then rely on a linear perturbation to minimize the applied work for finite-time protocols~\cite{Sivak2012, Zulkowski2012, Zulkowski2015}. An explicit solution has been derived for a passive Brownian particle in a harmonic confinement~\cite{Seifert2007, Schmiedl2008, Then2008, Zulkowski2013}. In general, the optimization depends on the complex interplay between the protocol details and the tracer relaxation within the confining potential~\cite{Sekimoto1997, Marcus2014}. Considering an equilibrium bath, the maximum power extracted from cyclic protocols can be expressed in terms of forces and associated fluxes within linear thermodynamics close to equilibrium: this approach does not need any reference to microscopic details~\cite{Seifert2008, Esposito2010, Proesmans2016}. However, for an actual colloidal engine which typically operates far from the linear regime, anticipating the properties of the maximum power is still largely a challenge to physical intuition.

Self-propelled particles extract energy from their environment to produce a sustained directed motion~\cite{Marchetti2013, Marchetti2018}. Canonical examples at the microscale are given by biological swimmers, such as bacteria~\cite{Elgeti2015}, and artificial colloidal particles, such as Janus colloids in a fuel bath~\cite{Bechinger2016}. Recent progress in the design of synthetic self-propelled colloids have established the possibility to control externally their activity~\cite{Palacci2013, Bechinger2013}. Besides, it has been shown recently that the activity of bacteria could also be modulated by some specific illuminations of the system~\cite{Walter2007, Leonardo2017, Arlt2018, Leonardo2018}. These setups motivate the search for new protocols, based on manipulating a self-propelled tracer instead of a passive one and, possibly, tuning periodically its swimming properties. It remains to determine whether such active colloidal engines could outperform their passive counterparts.


In this Letter, we consider colloidal engines operating with an overdamped self-propelled tracer immersed in a bath of passive Brownian particles. For simplicity, we model the trap applied by the external operator with a harmonic potential $ U_\text{ext} = \kappa {\bf x}^2 / 2 $ whose variance $\kappa$ can vary in time. This mimics the effect of the optical tweezers commonly used in actual experiments~\cite{Blickle2011, Martinez2016}. Introducing the tracer-bath and bath-bath interaction potentials, respectively denoted by $\sum_i U({\bf r}_i-{\bf x})$ and $\sum_{i<j} V ({\bf r}_i-{\bf r}_j)$, the dynamics of the system \{tracer+bath\} follows as
\begin{equation}\label{eq:dyn}
	\begin{aligned}
		\dot {\bf x} &= - \kappa {\bf x} - \nabla_{\bf x} \sum_i U ({\bf r}_i-{\bf x}) + {\bf f} ,
		\\
		\dot {\bf r}_i &= - \mu \nabla_i \Big[ \sum_j V({\bf r}_i-{\bf r}_j) + U ({\bf r}_i-{\bf x}) \Big] + {\boldsymbol\eta}_i ,
		\\
	\end{aligned}
\end{equation}
where we have set the tracer mobility to unity, and $\mu$ refers to the bath mobility. The thermal noise $\{{\boldsymbol\eta}_i\}$ is taken as zero-mean Gaussian with correlations $\langle \eta_{i\alpha}(t) \eta_{j\beta}(0) \rangle = 2 \mu T \delta_{ij} \delta_{\alpha\beta} \delta(t) $, where $T$ denotes the bath temperature. The Latin and Greek indices respectively refer to the particle label and to the spatial component. Following recent works~\cite{Szamel2015, Maggi2015, Brader2015, Nardini2016}, we model the self-propulsion force $\bf f$ as another zero-mean Gaussian noise, uncorrelated with $\{{\boldsymbol\eta}_i\}$ and with autocorrelation $ \langle f_\alpha (t) f_\beta(0) \rangle = \delta_{\alpha\beta}(T/\tau) \ee^{ - |t| / \tau } $. The persistence time $\tau$ embodies the typical time needed for the self-propulsion orientation to relax.

The damping being instantaneous while the self-propulsion correlations contain some memory, the system operates far from equilibrium. Note that the amplitude of the self-propulsion fluctuations are scaled such as the limit of vanishing persistence amounts to considering a passive tracer at temperature $T$: $ \langle f_\alpha (t) f_\beta(0) \rangle \underset{\tau\to0}{\longrightarrow} 2 T \delta_{\alpha\beta} \delta(t)$, in which case we recover an equilibrium dynamics for the system \{tracer+bath\}. Besides, the microscopic conversion of chemical energy into mechanical work is not explicitly described in~\eqref{eq:dyn}, but only considered implicitly in the resulting tracer self-propulsion.

In the following, we introduce two types of engines: Engine A operating with cycles of temperature $T$ and trap stiffness $\kappa$ at fixed tracer persistence $\tau$, and Engine B operating with cycles of tracer persistence $\tau$ and trap stiffness $\kappa$ at fixed temperature $T$. As a prelude, using techniques introduced in~\cite{Demery2011, Demery2014}, we trace out the bath degrees of freedom within some adiabatic mean-field treatment to reduce the system dynamics into an effective tracer dynamics. Such a treatment allows us to derive and optimize the output power of the engines in terms of the microscopic details. At first sight, one might expect that tracer activity generically leads to increase the performances of engines with respect to their passive counterparts. Yet, we reveal that Engine A, working at fixed activity, can only achieve less maximum power than its passive limit at vanishing persistence. In contrast, we also show that cyclic variation of activity in Engine B, which operates under isothermal condition without any passive equivalent, now allows one to find an optimal power in terms of tracer persistence and trap stiffness.


As a first step, we integrate the dynamics of the passive elements of the bath to only characterize effectively the motion of the tracer particle. To describe the bath dynamics from a coarse-grained viewpoint, we introduce the bath density $ \rho ({\bf r}, t) = \sum_i \delta [ {\bf r} - {\bf r}_i (t) ] $ and derive its time-evolution with standard techniques~\cite{Dean1996}:
\begin{equation}\label{eq:rho}
	\begin{aligned}
	\partial_t \rho ({\bf r}, t) &= \mu \nabla_{\bf r} \cdot \bigg[ \rho({\bf r}, t) \int \rho({\bf r}', t) \nabla_{\bf r} V ({\bf r}-{\bf r}') \dd{\bf r}' \bigg]
	\\
	&\quad + \mu \nabla_{\bf r} \cdot \Big[ \rho({\bf r}, t) \nabla_{\bf r} U ({\bf r}-{\bf x}) + T \nabla_{\bf r} \rho({\bf r}, t) \Big]
	\\
	&\quad + \nabla_{\bf r} \cdot \Big[ \sqrt{2 \mu \rho({\bf r}, t) T} {\boldsymbol\Lambda}({\bf r}, t) \Big] .
	\end{aligned}
\end{equation}
The fluctuating term $\boldsymbol\Lambda$ is a zero-mean Gaussian noise with correlations $ \langle \Lambda_\alpha ({\bf r}, t) \Lambda_\beta ({\bf r'}, t') \rangle = \delta_{\alpha\beta} \delta({\bf r}-{\bf r}') \delta(t-t') $.

To proceed further, we assume that the interactions among the particles composing the bath and those among the tracer and the bath particles are both weak. In such a regime, the system remains nearly homogeneous, so that the bath dynamics is only dictated by density fluctuations $\delta\rho = \rho - \rho_0$ around the average density $\rho_0$. It is thus natural to linearize~\eqref{eq:rho} around $\rho_0$ assuming $\mathcal{O}(U \delta \rho)\ll \mathcal{O}(\delta \rho)$. This amounts to considering perturbation around the mean-field limit at weak interactions~\cite{Campa2009, Bouchet2010, Nardini2012} and high density, namely for a dense bath of soft colloids~\cite{Demery2011, Demery2014}, where kinetic theories are known to work with high accuracy. The density dynamics is then solved in terms of Fourier components:
\begin{equation}\label{eq:delta-rho-c}
	\begin{aligned}
		\delta\rho_{\bf k} (t) = &\int_{-\infty}^t \ee^{ - \mu {\bf k}^2 ( T + \rho_0 V_{\bf k} ) (t-s) }
		\\
		& \times \Big[ \sqrt{2 \mu \rho_0 T} \ii {\bf k} \cdot {\boldsymbol\Lambda}_{\bf k}(s) - \mu {\bf k}^2 \rho_0 U_{\bf k} \ee^{\ii {\bf k}\cdot{\bf x}(s)} \Big] \dd s ,
	\end{aligned}
\end{equation}
where we have introduced the density mode as $ \delta\rho_{\bf k} (t) = \int \delta\rho({\bf r}, t) \ee^{\ii {\bf k}\cdot{\bf r}} \dd {\bf r} $. We defer the detailed derivation to~\cite{Supplemental}.

To obtain the effective tracer dynamics, we compute the force exerted by the bath on the tracer, which is expressed in terms of the density fluctuations as
\begin{equation}\label{eq:force-tracer-bath}
	- \nabla_{\bf x} \sum_i U ({\bf r}_i-{\bf x}) = \int \ii {\bf k} U_{\bf k} \delta\rho_{\bf k} \ee^{-\ii {\bf k}\cdot{\bf x}} \frac{\dd {\bf k}}{(2\pi)^d} ,
\end{equation}
where $d$ refers to the spatial dimension. As detailed in~\cite{Demery2011, Demery2014}, the force~\eqref{eq:force-tracer-bath} can be decomposed into a damping term and a zero-mean Gaussian noise. The former embodies the effect of the tracer in the surrounding bath, which in turn resists the tracer motion, while the latter reflects the effect of the bath noise on the tracer dynamics. The noise correlations generally contain some memory which depends on the tracer position. However, memory effects become irrelevant when the bath relaxation around the tracer, controlled by the diffusive time $\tau_\text{diff} = \sigma^2 / \mu ( T + \rho_0 V_{|{\bf k}|=0} )$ on the tracer scale $\sigma$, is much faster than the tracer relaxation in the trap, which takes a typical time $\kappa^{-1}$.

In the adiabatic limit $ \kappa \tau_\text{diff} \ll 1 $, the effective tracer dynamics then amounts to the following Langevin equation
\begin{equation}\label{eq:tracer}
	( 1 + \lambda ) \dot {\bf x} = - \kappa {\bf x} + {\bf f} + {\boldsymbol\xi} ,
\end{equation}
where $\boldsymbol\xi$ is a zero-mean Gaussian noise, uncorrelated with $\bf f$, with correlations $\langle \xi_\alpha(t) \xi_\beta(0) \rangle = 2 \lambda T \delta_{\alpha\beta} \delta (t)$. The dimensionless coefficient $\lambda $ captures the effect of interactions with surrounding particles. It can be expressed in terms of the microscopic details of the bath for generic interactions. Neglecting the size of bath particles when interacting between each other: $ V_{\bf k} \simeq V_{|{\bf k}|=0} $, and introducing the bath-tracer energetic scale $U_0$ such as $ U({\bf x}) = U_0 \phi({\bf x}) $, we evaluate $\lambda$ as~\cite{Supplemental}
\begin{equation}\label{eq:lambda}
	\lambda = \frac{\rho_0}{\mu d} \bigg[ \frac{U_0}{T + \rho_0 V_{|{\bf k}|=0}} \bigg]^2 \int [\phi({\bf x})]^2 \dd {\bf x} .
\end{equation}
In short, the adiabatic mean-field treatment, valid for weak interactions and high density at $\kappa \tau_\text{diff} \ll 1$, allows us to reduce the original dynamics~\eqref{eq:dyn} into the effective dynamics~(\ref{eq:tracer}--\ref{eq:lambda}) for the tracer only: the effect of interactions with surrounding particles is analogue to coupling with a thermal bath at temperature $T$ with drag coefficient $\lambda$. Note that such an effective dynamics does not rely on any response theory, at variance with some recent works~\cite{Steffenoni2016, Maes2017, England2018}, thus allowing one to formulate it explicitly for given interactions.

\begin{figure}
	\centering
	\includegraphics[width=1.0\columnwidth]{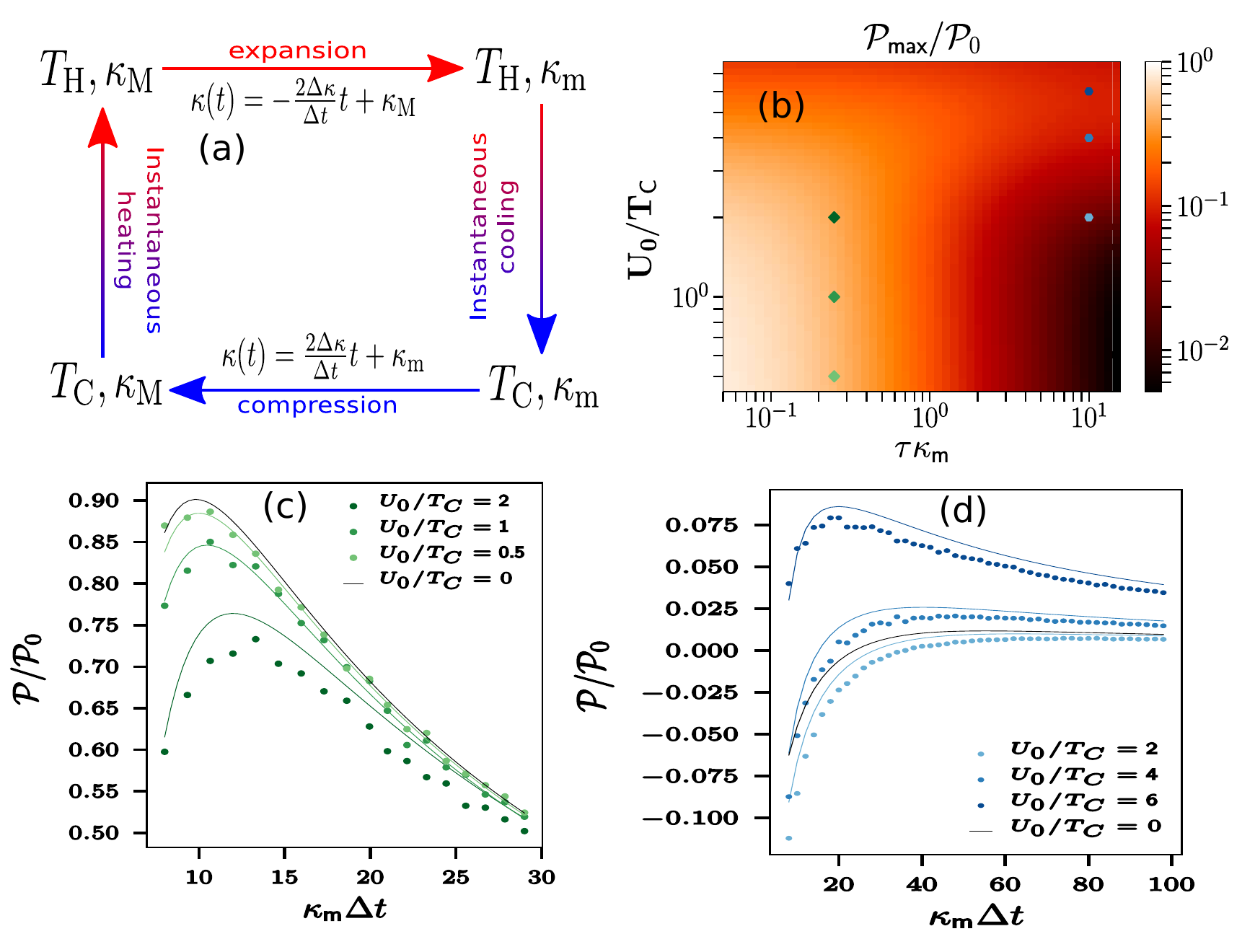}
	\caption{\label{fig:engA}
		(Color online)
		(a)~Schematic representation of Engine A: cyclic variation of temperature $T$ and trap stiffness $\kappa$ at fixed tracer persistence $\tau$.
		(b)~Maximum power ${\cal P}_\text{max}$ scaled by the value for a non-interacting passive tracer ${\cal P}_0$ as a function of the scaled coupling with the bath $U_0/T_\text{\tiny C}$ and of the scaled persistence $\kappa_\text{m}\tau$. Each point in the color map is obtained by first integrating~\eqref{eq:moments} for different $\Delta t$ and then optimizing to extract the maximum power.
		(c-d)~Scaled power ${\cal P}/{\cal P}_0$ as a function of the scaled cycle time $\kappa_\text{m}\Delta t$ for the parameter values referred to by colored markers in (b): simulations of the microscopic dynamics~\eqref{eq:dyn} and analytic results obtained by integrating~\eqref{eq:moments} are respectively in dots and solid lines. Simulation details and parameter values are given in~\cite{Supplemental}.
	}
\end{figure}

The first engine that we consider is the analogue of the colloidal Stirling engine~\cite{Blickle2011, Schwieger2015}, except that the tracer is now a self-propelled particle. Engine A operates with four successive branches described in Fig.~\ref{fig:engA}(a): (i) the operator compresses the trap from $\kappa_\text{m}$ to $\kappa_\text{\tiny M}$ at low temperature $T_\text{\tiny C}$, (ii) the system is heated up from $T_\text{\tiny C}$ to $T_\text{\tiny H}$ at high trap stiffness $\kappa_\text{\tiny M}$, (iii) the operator expands the trap from $\kappa_\text{\tiny M}$ to $\kappa_\text{m}$ at high temperature $T_\text{\tiny H}$, and (iv) the system is cooled down from $T_\text{\tiny H}$ to $T_\text{\tiny C}$ at small trap stiffness $\kappa_\text{m}$.

Following stochastic thermodynamics~\cite{Sekimoto1998, Seifert2012}, the stochastic work exerted on the particles under external modification of the confinement potential $U_\text{ext}$ during a given protocol time $\Delta t$  reads $\int_0^{\Delta t} \partial_t U_\text{ext} \dd t$. Importantly, such a definition does not depend on the details of the bath, nor on the nature of the tracer, either passive or active, since it only concerns the interaction between the tracer and the external operator. In particular, it is independent of the presence of any dissipation of energy in the thermostat, characteristic of active systems~\cite{Toyabe2010, Ahmed2016, Nardini2016, Speck2016, Mandal2017, Seifert2018, Shankar2018}. Indeed, a given realization of the work is entirely determined by the particle trajectory during the protocol, so that similar trajectories for different tracers should lead to the same extracted work. Of course, the explicit evaluation of the work statistics will depend on the activity of the tracer and on the interactions with surrounding particles. Note that it would differ if the tracer dynamics~\eqref{eq:dyn} contained some exponential memory kernel in the damping force, with same coefficients as the correlations of $\bf f$, in which case the dynamics would operate in equilibrium without any dissipation.

For the harmonic trap $U_\text{ext} = \kappa {\bf x}^2 / 2$, the average work then reads $ {\cal W} = (1/2) \int_0^{\Delta t} \dot \kappa \langle {\bf x}^2 \rangle \dd t $. To obtain explicitly the average work as a function of the microscopic parameters, one needs to derive the time-evolution of the tracer variance. Its dynamics is coupled to the one of $\langle {\bf x}\cdot{\bf f} \rangle$ and $\langle {\bf f}^2 \rangle$ in a closed formed given by~\cite{Supplemental}
\begin{equation}\label{eq:moments}
	\begin{aligned}
		\frac{\dd \langle {\bf x}^2 \rangle}{\dd t} &= - \frac{2 \kappa \langle {\bf x}^2 \rangle}{1+\lambda} + \frac{2 \langle {\bf x}\cdot{\bf f} \rangle}{1+\lambda} + \frac{2 d \lambda T}{(1+\lambda)^2} ,
		\\
		\frac{\dd \langle {\bf x}\cdot{\bf f} \rangle}{\dd t} &= - \left[ \frac{\kappa}{1+\lambda} + \frac{1}{\tau} \right] \langle {\bf x}\cdot{\bf f} \rangle + \frac{\langle {\bf f}^2 \rangle}{1+\lambda} ,
		\\
		\frac{\dd \langle {\bf f}^2 \rangle}{\dd t} &= - \frac{2 \langle {\bf f}^{2} \rangle}{\tau} + \frac{2 d T}{\tau^2} .
	\end{aligned}
\end{equation}
As a first insight into the performances of Engine A, we examine a quasistatic protocol for which the tracer statistics fully relaxes between two successive infinitesimal variations of the trap stiffness. The corresponding work is given by $ {\cal W}_\infty = (1/2) \int \langle {\bf x}^2 \rangle_\text{\tiny S} \dd \kappa $, where $\langle {\bf x}^2 \rangle_\text{\tiny S}$ refers to the stationary tracer variance, which can be obtained from~\eqref{eq:moments} as $ \langle {\bf x}^2 \rangle_\text{\tiny S} = (d T/\kappa) [ 1 - \kappa\tau / (1+\lambda) (1+\lambda+\kappa\tau) ] $. The quasistatic work of Engine A follows as
\begin{equation}
	\begin{aligned}
		{\cal W}_\text{\tiny A} &= \frac{d ( T_\text{\tiny C} - T_\text{\tiny H} )}{2} \ln \left[ \frac{ \kappa_\text{\tiny M} }{ \kappa_\text{m} } \right] - \frac{ d T_\text{\tiny C} / 2 }{ 1 + \lambda_\text{\tiny C} } \ln \left[ \frac{1 + \lambda_\text{\tiny C} + \kappa_\text{\tiny M}\tau}{1 + \lambda_\text{\tiny C} + \kappa_\text{m}\tau} \right]
		\\
		& \quad + \frac{ d T_\text{\tiny H} / 2 }{ 1 + \lambda_\text{\tiny H} } \ln \left[ \frac{1 + \lambda_\text{\tiny H} + \kappa_\text{\tiny M}\tau}{1 + \lambda_\text{\tiny H} + \kappa_\text{m}\tau} \right] ,
		\end{aligned}
\end{equation}
where $ \lambda_\text{\tiny C} = \lambda (T_\text{\tiny C})$ and  $ \lambda_\text{\tiny H} = \lambda (T_\text{\tiny H})$. Note that ${\cal W}_\text{\tiny A}$ is always negative: Engine A extracts some work from a quasistatic protocol for all values of the tracer persistence and tracer-bath interactions.

The limits of small and large persistence times $\tau$ can be rationalized with simple physical arguments. For a vanishing persistence, the system \{tracer+bath\} is at equilibrium. Then, the stationary tracer variance is given by the equipartition theorem: $\langle {\bf x}^2 \rangle_\text{\tiny S} = d T / \kappa$, so that the quasistatic work is independent of the interaction parameter $\lambda$ and reads $ {\cal W}_\text{\tiny A} \underset{\tau\to0}{\longrightarrow} (d/2) (T_\text{\tiny C} - T_\text{\tiny H} ) \ln ( \kappa_\text{\tiny M} / \kappa_\text{m} ) $. Instead, at large persistence, the self-propulsion becomes deterministic, so that the only source of fluctuations in the tracer dynamics arises from interactions with the surrounding bath particles. The corresponding extracted work can be expressed as $ {\cal W}_\text{\tiny A} \underset{\tau\to\infty}{\longrightarrow} (d/2) [ \lambda_\text{\tiny C} T_\text{\tiny C} / (1+\lambda_\text{\tiny C}) - \lambda_\text{\tiny H} T_\text{\tiny H} / (1+\lambda_\text{\tiny H}) ] \ln ( \kappa_\text{\tiny M} / \kappa_\text{m} ) $. It vanishes in the absence of interactions, {\it i.e.} for $\lambda=0$, since the tracer is no longer subject to any fluctuations in such a limit.

We now turn to discuss how the tracer activity and interactions with the bath affect the finite-time properties. For simplicity, and taking inspiration from protocols that have been used in experiments~\cite{Blickle2011}, we consider instantaneous temperature changes and linear variations of the trap stiffness in time, with equal duration $\Delta t /2$ for compression and expansion, as depicted in Fig.~\ref{fig:engA}(a). Owing to the linearity of the coupled dynamics~\eqref{eq:moments}, one can derive the tracer variance as a function of time in the compression and expansion branches separately. Each solution is parametrized by the initial value at the beginning of the branch. In steady state, the initial compression value $\langle {\bf x}^2 (0) \rangle$ should coincide with the final expansion value $\langle {\bf x}^2 (\Delta t) \rangle$, and similarly the final compression value and the initial expansion value should both be equal to $\langle {\bf x}^2 (\Delta t/2) \rangle$, since the temperature changes are instantaneous. Such a constraint is enforced by a set of fixed point equations in the dynamics of $\{ \langle {\bf x}^2 \rangle, \langle {\bf x}\cdot{\bf f} \rangle, \langle {\bf f}^2 \rangle \}$~\cite{Supplemental}. Solving these equations for different cycle times $\Delta t$ allows us to obtain the average work as a function of $\Delta t$.

We find that the extracted work decreases monotonically with the cycle time, and it becomes positive at short times: when operated too rapidly, {\it i.e.} when the relaxation of the tracer in the trap can no longer follow the external drive, Engine A does not extract work from the bath. This qualitative feature is known to be generic, and it has been already captured within linear response for arbitrary engines operating with an equilibrium bath~\cite{Seifert2008, Esposito2010}. More information can be extracted from the output power ${\cal P} = - {\cal W} / \Delta t$. It vanishes for large cycle times, since the work remains finite in the quasistatic limit, and it has typically a maximum value ${\cal P}_\text{max}$ for a finite cycle time, which reflects the trade-off between fast external driving and slow tracer relaxation. We focus on the dependence of ${\cal P}_\text{max}$ in terms of the bath-tracer interactions and the tracer persistence to characterize the engine performance.

At fixed interactions, {\it i.e.} when $U_0/T$ or analogously $\lambda$ is constant, the maximum power always decreases with the tracer persistence $\tau$: the tracer activity is generically a drawback for Engine A. Instead, at fixed persistence, increasing the interactions can lead to two opposite effects depending on the ratio of the persistence time to the relaxation time within the trap: (i) when $\tau \kappa_\text{m} \ll 1$, the system is close to equilibrium and we observe that increasing the coupling $\lambda$ of the tracer with the bath, which is analogue to increasing the solvent drag coefficient in~\eqref{eq:tracer}, decreases ${\cal P}_\text{max}$; (ii) when $\tau \kappa_\text{m} \gg 1$, the self-propulsion force is almost deterministic and we find that increasing such a coupling, which now amounts to thermalizing the tracer by progressively neglecting the self-propulsion, increases ${\cal P}_\text{max}$, as reported in Fig.~\ref{fig:engA}(b).

It is important to realize that these predictions rely on our adiabatic mean-field treatment of the bath, which, in particular, should only be valid for weak interactions {\it a priori}~\cite{Demery2011, Demery2014}. To investigate the range of validity of our approach, we compare the output power as a function of cycle time obtained (i) by solving~\eqref{eq:moments} for different $\Delta t$ as described above, and (ii) from direct numerical simulations of the microscopic dynamics~\eqref{eq:dyn}. We consider bath particles in two dimensions which interact {\it via} a short-range repulsion of the form $ V({\bf r}) = \varepsilon (1-r/a)^2 \Theta(a-r) $, where $\Theta$ refers to the Heaviside step function, and we account for the bath-tracer interactions through a Gaussian potential: $ U({\bf x}) = U_0 \ee^{ - ({\bf x}/\sigma)^2 / 2 } $~\cite{Supplemental}. We observe a quantitative agreement between numerics and predictions for moderate interactions, as expected; see light green and light blue curves in Figs.~\ref{fig:engA}(c,d). Some deviations become manifest as the coupling to the bath is enhanced, yet the qualitative trend remains similar: the maximum power decreases (increases) at small (large) persistence, as reported for the darker blue and darker green curves in Figs.~\ref{fig:engA}(c,d). As a result, direct numerical simulations support the validity of our approach even beyond the mean-field regime.


To go beyond the protocols used in actual colloidal heat engines~\cite{Blickle2011, Martinez2016, Sood2016}, which are commonly inspired by the ones of macroscopic engines operating with a thermal bath, the activity of the tracer can be regarded as an additional control parameter to be tuned externally. The second law of thermodynamics enforces that any isothermal heat engine cannot operate with an equilibrium bath. Work can only be extracted from the energy flow induced by a temperature difference~\cite{Carnot, Callen}. In contrast, the energy dissipated by the self-propelled particle while moving persistently in the solvent induces a steady flux of energy into the thermostat at constant temperature~\cite{Nardini2016, Speck2016, Mandal2017, Seifert2018, Shankar2018}. It follows that isothermal heat engines can generically be designed when manipulating an active tracer instead of a passive one. In our settings, this is manifest by considering an engine based on varying tracer persistence and trap stiffness.

\begin{figure}
	\centering
	\includegraphics[width=1.0\columnwidth]{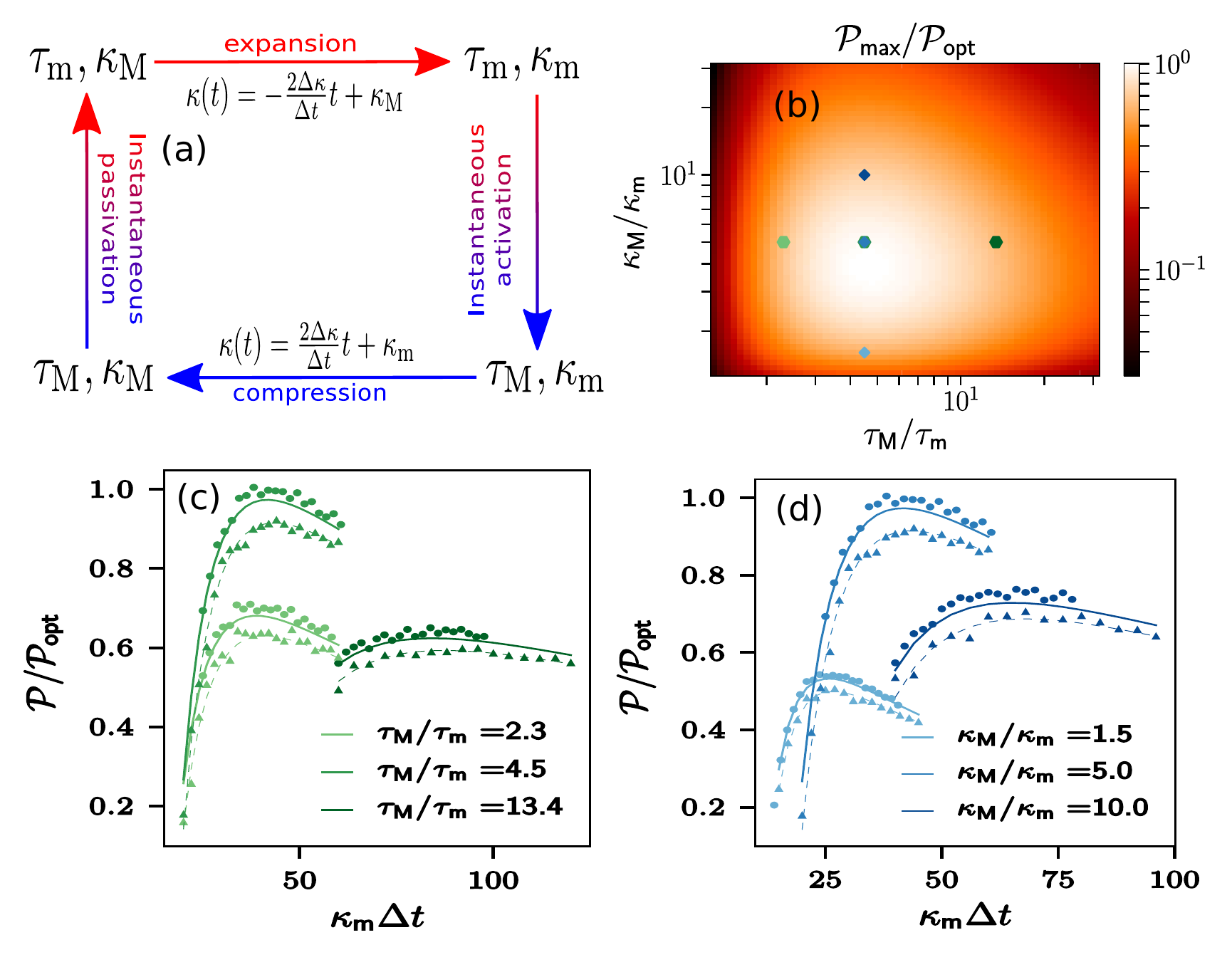}
	\caption{\label{fig:engB}
		(Color online)
		(a)~Schematic representation of Engine B: cyclic variation of tracer persistence $\tau$ and trap stiffness $\kappa$ at fixed temperature $T$.
		(b)~Maximum power ${\cal P}_\text{max}$ scaled by the optimal value ${\cal P}_\text{opt}$ as a function of the ratio between the maximum and the minimum values of stiffnesses $\kappa_\text{\tiny M}/\kappa_\text{m}$ and of persistences $\tau_\text{\tiny M}/\tau_\text{m}$, in the absence of coupling with the bath.
		(c-d)~Scaled power ${\cal P}/{\cal P}_\text{opt}$ as a function of the scaled cycle time $\kappa_\text{m}\Delta t$ for the parameter values referred to by colored markers in (b): solid and dashed lines refer to analytical results obtained by integrating~\eqref{eq:moments} respectively without ($\lambda=0$) and with ($\lambda\neq0$) interactions with the bath; circles and triangles refer to simulations of~\eqref{eq:dyn} respectively without ($U=0$) and with ($U\neq0)$ interactions with the bath. Simulation details and parameter values are given in~\cite{Supplemental}.
	}
\end{figure}

Engine B consists in replacing the heating (cooling) branch in Engine A by some passivation (activation) branch, for which the tracer persistence decreases (increases) at fixed temperature and trap stiffness, as described in Fig.~\ref{fig:engB}(a). The quasistatic work can be deduced from the stationary tracer variance as
\begin{equation}
	{\cal W}_\text{\tiny B} = - \frac{dT/2}{1+\lambda} \ln \left[ \frac{1 + \lambda + \kappa_\text{\tiny M}\tau_\text{\tiny M}}{1 + \lambda + \kappa_\text{\tiny m}\tau_\text{\tiny M}} \cdot \frac{1 + \lambda + \kappa_\text{\tiny m}\tau_\text{\tiny m}}{1 + \lambda + \kappa_\text{\tiny M}\tau_\text{\tiny m}} \right] ,
\end{equation}
where $\tau_\text{\tiny M}$ and $\tau_\text{\tiny m}$ respectively refer to the maximum and minimum persistence values. Note that such a work does not account for the implicit protocols modifying tracer persistence, since the microscopic mechanism at the basis of self-propulsion is not explicit in the dynamics~\eqref{eq:dyn}.

At variance with Engine A, Engine B extracts work only out of the self-propulsion fluctuations, since no work is extracted from the thermal fluctuations, stemming from interactions with bath particles, at fixed temperature. Then, the effect of increasing interactions is always to reduce the quasistatic work, by progressively overwhelming self-propulsion fluctuations. In particular, for a strong coupling $\lambda\gg1$, the tracer activity becomes completely irrelevant and thus the extracted work vanishes. Moreover, the limits of small and large persistence correspond respectively to (i) mapping self-propulsion fluctuations into thermal ones, and (ii) neglecting self-propulsion fluctuations. Work cannot be extracted in both regimes, so that ${\cal W}_\text{\tiny B}$ vanishes in each limit.

For finite cycle times, we consider compression and expansion branches operating with linear variations of trap stiffness, by analogy with Engine A, while the persistence jumps instantaneously in between these branches, as shown in Fig.~\ref{fig:engB}(a). We apply the same procedure as for Engine A to obtain the output power as a function of the cycle time~\cite{Supplemental}. The maximum power ${\cal P}_\text{max}$ decreases monotonically with the coupling to the bath $\lambda$, as the extracted work does; see Figs.~S1(a,b) of~\cite{Supplemental}. Interestingly, there is an optimum of ${\cal P}_\text{max}$ in terms of the persistence ratio $\tau_\text{\tiny M} / \tau_\text{m}$ and of the trap stiffness ratio $\kappa_\text{\tiny M} / \kappa_\text{m}$, as reported in Fig.~\ref{fig:engB}(b). The optimal value increases linearly with temperature $T$. The power optimum emerges from a trade-off between (i) large enough persistence and stiffness ratios to extract some power, and (ii) small enough persistence and stiffness so that the self-propulsion fluctuations are still relevant in the tracer dynamics. This is in contrast with Engine A for which ${\cal P}_\text{max}$ increases monotonously with the ratio of both temperatures and trap stiffnesses, as shown in Fig.~S1(c) of~\cite{Supplemental}.

To confirm these predictions, we extract the maximum power from direct simulations of the microscopic dynamics~\eqref{eq:dyn}, for different values of persistence and stiffness ratios across the optimal set of parameters shown in Fig.~\ref{fig:engB}(b). The numerics are in close agreement with analytical results for each one of these parameter values. They support that the power is close to optimal for the chosen set of parameters, as reported in Figs.~\ref{fig:engB}(c,d). Besides, including interactions systematically reduces the power as a function of cycle time, as expected analytically.


In this Letter, we have introduced and studied two minimal heat engines built out of a self-propelled particle immersed in a thermal bath formed by passive colloids. For a non-isothermal engine with time-independent activity, the self-propulsion generically reduces the maximum available power, whereas increasing the coupling with the surrounding bath compensates for such an effect by thermalizing the tracer. However, one can take advantage of the tracer activity to design an {\it isothermal engine}, based on synchronizing the tracer persistence with the trap stiffness, whose maximum power is optimal for a given set of persistence and stiffness ratios. These results address regimes which are beyond the linear perturbation with respect to quasistatic protocols~\cite{Seifert2007, Sivak2012}, since the engines operate far from any steady state of the system at maximum power. Our approach, relying on an adiabatic mean-field treatment of the bath, is quantitatively valid when interactions are weak~\cite{Demery2011, Demery2014}. Direct simulations of the microscopic dynamics support that it remains qualitatively robust even beyond the regime where our theory is mathematically well funded. These analytical techniques could be extended to investigate the maximum power of various engines, either for an equilibrium or a nonequilibrium bath, operating with more complex protocols. More generally, it opens the door to predicting the finite-time properties of heat engines beyond the colloidal case~\cite{Steeneken2011, Sothmann2012, Lutz2014, Lutz2016}. Finally, the efficiency of heat engines, defined as the ratio of extracted work to dissipated heat, also generally arises as a natural way to evaluate the performances of cyclic protocols. Yet, for self-propelled particles, evaluating properly the amount of dissipated energy is still largely subject to interpretation~\cite{Speck2016, Speck2017, Mandal2017, Seifert2018}. Then, we defer the study of efficiency in active colloidal heat engines to future works, with the hope to elucidate its relation to optimal power, by analogy with standard heat engines~\cite{Seifert2008, Esposito2009, Allahverdyan2013, Shiraishi2016, Pietzonka2017}.


\acknowledgements{
	The authors acknowledge insightful discussions with Vincent D\'emery, Julien Tailleur and Fr\'ed\'eric van Wijland. \'EF benefits from an Oppenheimer Research Fellowship from the University of Cambridge, as well as a Junior Research Fellowship from St Catherine's College. CN acknowledges the support of an Aide Investissements d'Avenir du LabEx PALM (ANR-10-LABX-0039-PALM). Work funded in part by the ERC under the EU Horizon 2020 Programme via ERC grant agreement 740269.
}


\bibliographystyle{eplbib}
\bibliography{ActiveEngine_ref}

\end{document}